# CORONAL MASS EJECTIONS AND THE SOLAR CYCLE VARIATION OF THE SUN'S OPEN FLUX

Y.-M. Wang and N. R. Sheeley, Jr.  
Space Science Division, Naval Research Laboratory, Washington, DC 20375, USA; yi.wang@nrl.navy.mil, neil.sheeley@nrl.navy.mil

## ABSTRACT

The strength of the radial component of the interplanetary magnetic field (IMF), which is a measure of the Sun's total open flux, is observed to vary by roughly a factor of two over the 11 year solar cycle. Several recent studies have proposed that the Sun's open flux consists of a constant or "floor" component that dominates at sunspot minimum, and a time-varying component due to coronal mass ejections (CMEs). Here, we point out that CMEs cannot account for the large peaks in the IMF strength which occurred in 2003 and late 2014, and which coincided with peaks in the Sun's equatorial dipole moment. We also show that near-Earth interplanetary CMEs, as identified in the catalog of Richardson and Cane, contribute at most ∼30% of the average radial IMF strength even during sunspot maximum. We conclude that the long-term variation of the radial IMF strength is determined mainly by the Sun's total dipole moment, with the quadrupole moment and CMEs providing an additional boost near sunspot maximum. Most of the open flux is rooted in coronal holes, whose solar cycle evolution in turn reflects that of the Sun's lowest-order multipoles.

## 1. INTRODUCTION

The radial component of the interplanetary magnetic field (IMF) originates from the small fraction of the flux threading the solar surface which is dragged into the heliosphere by the solar wind and by coronal mass ejections (CMEs). *Ulysses* magnetometer measurements have shown that the radial IMF strength, $|B_r|$, is approximately independent of heliographic latitude $L$ (Balogh et al. 1995; Smith & Balogh 2008; Erdős & Balogh 2014). It then follows that $|B_r|$ (averaged over longitude $\phi$) is related to the Sun's total unsigned open flux, $\Phi_{\rm open}$, by

$$|B_r(r)| = \frac{\Phi_{\rm open}}{4\pi r^2} \qquad (r \gg R_\odot), \qquad (1)$$

where $r$ is heliocentric distance. Near-Earth spacecraft measurements[1] indicate that $|B_r|$ varies by a factor of the order of two over the 11 year sunspot cycle, with the highest peaks in cycles 21–23 occurring ∼2 years after the sunspot number attained its maximum. This Letter addresses the question of what causes the solar cycle variation of $|B_r|$, or equivalently, of $\Phi_{\rm open}$.

If the contribution of CMEs is ignored, $\Phi_{\rm open}$ can be estimated by applying a potential-field source-surface (PFSS) extrapolation to Carrington maps of the observed photospheric field. Here, the coronal field $\boldsymbol{B}(r, L, \phi)$ is assumed to remain current-free out to a spherical source surface $r = R_{\rm ss}$, where the field lines are constrained to be radial (Schatten et al. 1969); at the inner boundary $r = R_\odot$, $B_r$ is matched to the photospheric field, which is taken to be radially oriented at the depth where it is measured (Wang & Sheeley 1992). All field lines that cross the source surface are defined to be "open," with their footpoint areas representing coronal holes. The total open flux is then found by integrating $|B_r(R_{\rm ss}, L, \phi)|$ over the source surface. Setting $R_{\rm ss} \simeq 2.5\ R_\odot$ yields a surprisingly good match to observations of the interplanetary sector structure, streamer patterns, and coronal holes throughout the solar cycle (see, e.g., Hoeksema 1984; Wang et al. 1996).

In an earlier study (Wang & Sheeley 1995), we derived the time variation of $\Phi_{\rm open}$ by applying the PFSS model to Carrington maps from the Mount Wilson Observatory (MWO) and the Wilcox Solar Observatory (WSO), both of which employ the Zeeman-sensitive Fe I 525.0 nm line to measure the line-of-sight photospheric field. We found that we could reproduce the observed long-term variation of $B_{\rm E}$, the radial IMF strength at Earth, after correcting the measured photospheric fluxes for the effects of line profile saturation and flux-tube physics by multiplying them by the factor $(4.5-2.5 \sin^2 L)$. This latitude- (or center-to-limb angle) dependent correction was obtained by Ulrich (1992; see also Ulrich et al. 2002, 2009) by comparing MWO measurements made in Fe I 525.0 nm with those made in the nonsaturating line Fe I 523.3 nm. If instead we adopted the latitude-independent correction factor of 1.8 inferred by Svalgaard et al. (1978) from an analysis of WSO observations in the Fe I 525.0 nm line alone, we found that Equation (1) gave values of $B_{\rm E}$ that were a factor of ∼2 too small and whose variation was out of phase with the observed behavior. One possible source of the discrepancy between the two calibration corrections is the implicit assumption by Svalgaard et al. (1978) that the shape of the Fe I 525.0 nm line profile remains the same everywhere on the Sun, whereas in fact the profile becomes shallower in magnetic regions, leading to large underestimates of the actual flux if the magnetograph exit slit is positioned near the line core (Harvey & Livingston 1969).

In the PFSS model, the variation of the total open flux largely follows that of the Sun's total dipole strength, since the dipole (and, at sunspot maximum, the quadrupole) component of the photospheric field provides the dominant contribution to the source surface field. The solar cycle evolution of these low-order multipoles is in turn determined by the emergence of active regions and the transport of their fields over the photosphere. The open flux itself is rooted in coronal holes, which are embedded within the unipolar areas formed by these transport processes.

---
[1] See http://omniweb.gsfc.nasa.gov.

Over the last decade, several authors have proposed a very different picture for the solar cycle variation of the IMF (see, e.g., Reinard & Fisk 2004; Owens & Crooker 2006; Riley 2007; Schwadron et al. 2010; Owens et al. 2011; Owens & Lockwood 2012). In this scenario, the observed IMF variation is due almost entirely to CMEs. A related concept is that of a fixed "floor" in the IMF strength and total open flux, which was supposedly approached during the activity minimum of 2008–2009 (Svalgaard & Cliver 2007; Owens et al. 2008; Yermolaev et al. 2009; Cliver & Ling 2011). In most of these models, the time-varying component of the IMF and open flux is closely correlated with the CME rate, which in turn is known to be correlated with the sunspot number (see, e.g., Webb & Howard 1994); the floor is reached when the CME rate vanishes.

In the following sections, we show that the solar cycle variation of the radial IMF strength does not follow the CME rate, and that near-Earth interplanetary CMEs (ICMEs) are not the main contributor to $B_E$, even at activity maximum.

## 2. THE DIFFERING SOLAR-CYCLE VARIATIONS OF THE CME RATE AND IMF STRENGTH

The Large Angle and Spectrometric Coronagraph (LASCO) on the *Solar and Heliospheric Observatory* (*SOHO*) has been observing CMEs from 1996 to the present. The times and properties of individual events are listed in a number of catalogs, including CACTus (Robbrecht & Berghmans 2004),[2] which employs automated detection techniques and also includes events observed with the *STEREO* COR2A/B coronagraphs since 2007. In this section, we use the CACTus catalog to derive the time variation of the CME rate, $N_{CME}$, and compare it with the variation of the radial IMF strength at Earth.

As noted in Wang & Colaninno (2014), the doubling of the LASCO image cadence after 2010 August led to an artificial increase in the number of LASCO CMEs detected by CACTus, and to a divergence between the LASCO and COR2 counts. To correct for this effect, we replace the LASCO count rates after mid-2010 by an average of the COR2A and COR2B count rates, taking $N_{CME} = N_{LASCO}$ before 2010 August and $N_{CME} = (N_{COR2A} + N_{COR2B})/2$ during 2010–2014 August. From 2014 September onward, when *STEREO A* and *B* were behind the Sun and the COR instruments were switched off, we set $N_{CME} = \xi_{cola} N_{LASCO}$, where $\xi_{cola}$ is the ratio of the COR2A/B and LASCO count rates averaged over the interval 2010–2014 August. Because of the low and irregular cadence of the LASCO observations during the early part of the *SOHO* mission, we consider only the CME rates from 1999 March onward.

Figure 1(a) shows, as a function of time $t$ during 1999–2015, the daily CME rate $N_{CME}$, the near-Earth radial IMF strength $B_E^{obs}$, and the monthly mean sunspot number $R_I$. Here and in subsequent figures, 3-Carrington rotation (CR) or 82-day running means have been taken; $B_E^{obs}$ is calculated by taking the absolute value of the (signed) daily averages of $B_r$ provided by the OMNI database. It is obvious from Figure 1(a) that $N_{CME}(t)$ and $B_E^{obs}(t)$ behave quite differently: whereas the CME rate varies in phase with the sunspot number, with correlation coefficient $cc(N_{CME}, R_I) = 0.89$, the radial IMF strength attains its maximum at the start of the declining phase of the cycle, with $cc(B_E^{obs}, R_I)$ and $cc(B_E^{obs}, N_{CME})$ both being as low as 0.56.

Figure 1(b) compares the variation of the radial IMF strength with that of the Sun's total open flux $\Phi_{open}$, as predicted by applying a PFSS extrapolation to WSO photospheric field maps, after multiplying the magnetograph measurements by the Ulrich correction $(4.5 - 2.5 \sin^2 L)$. The WSO-derived open flux (which assumes a source surface at $2.5\,R_\odot$) has been converted into a radial field strength at 1 AU using Equation (1), and is hereafter designated as $B_E^{WSO}$. Also plotted are $R_I(t)$ and $D_{eq}(t)$, the equatorial dipole or $(l = 1, |m| = 1)$ component of the WSO photospheric field. The variation of $B_E^{WSO}$ bears a striking resemblance to that of $B_E^{obs}$ ($cc = 0.90$), with both curves showing large peaks in 2003 and near the end of 2014. The peaks in $B_E^{WSO}(t)$ in turn coincide with large increases in $D_{eq}(t)$. As discussed in Wang & Sheeley (2003) and Sheeley & Wang (2015), the Sun's equatorial dipole strength depends not only on the level of sunspot activity but also on its distribution in longitude, with increases occurring when active regions emerge with their east–west dipole moments in phase with each other or with the preexisting background field. Thus, even though the rate of flux emergence is greatest at sunspot maximum, $D_{eq}$ tends to reach its maximum strength somewhat later, when activity becomes concentrated in a smaller number of very strong active regions having a relatively nonuniform distribution in longitude.

According to Owens & Crooker (2006), the IMF strength is determined by a balance between flux injection by CMEs and interchange reconnection between the CME loops and open field lines. As is evident from their Figure 5, the predicted IMF variation then approximately follows the CME rate. Inspection of their Figure 3 shows that their best-fit simulations, which use the LASCO CME rate and assume a reconnection timescale of ~50 days, overestimate the observed IMF strength during 2000–2002 but underestimate it during 2003. This is consistent with our conclusion that the actual IMF variation differs significantly from that of the CME rate.

## 3. CONTRIBUTION OF ICMEs TO THE RADIAL IMF VARIATION

The saturation correction of 1.8 deduced by Svalgaard et al. (1978), when used to derive $\Phi_{open}$ from WSO or MWO magnetograph measurements, yields values of $B_E$ that are ~2–3 times too low around sunspot maximum (see Figure 2(a) in Wang & Sheeley 1995). However, by adding the contribution of near-Earth ICMEs to $\Phi_{open}^{(Sv)}$, Riley (2007) was able to match the observed radial IMF variation through most of cycles 21–23, except for a large discrepancy that remained during 2003–2006. To estimate $B_E^{ICME}$, the ICME contribution to the radial IMF strength averaged over a CR, Riley assumed

$$B_E^{ICME} \sim \langle B_{ICME}\rangle N_{ICME} \left(\frac{\tau_{ICME}}{\tau_{CR}}\right), \qquad (2)$$

where $\langle B_{ICME}\rangle$ is taken as the total field strength inside a typical ICME, $N_{ICME}$ is the number of ICMEs during a CR, $\tau_{ICME}$ is the duration of each event, and $\tau_{CR} = 27.3$ days. By setting $\langle B_{ICME}\rangle \sim 8$ nT, $N_{ICME} \sim 0.02 R_I$, and $\tau_{ICME} = 1.5$ days, Riley (2007) found that ICMEs could account for over one-half of the observed radial IMF strength at sunspot maximum. However, this is an overestimate, because it is the radial component of the ICME field, not its total field, that should be used in Equation (2).

---

[2] http://sidc.oma.be/cactus

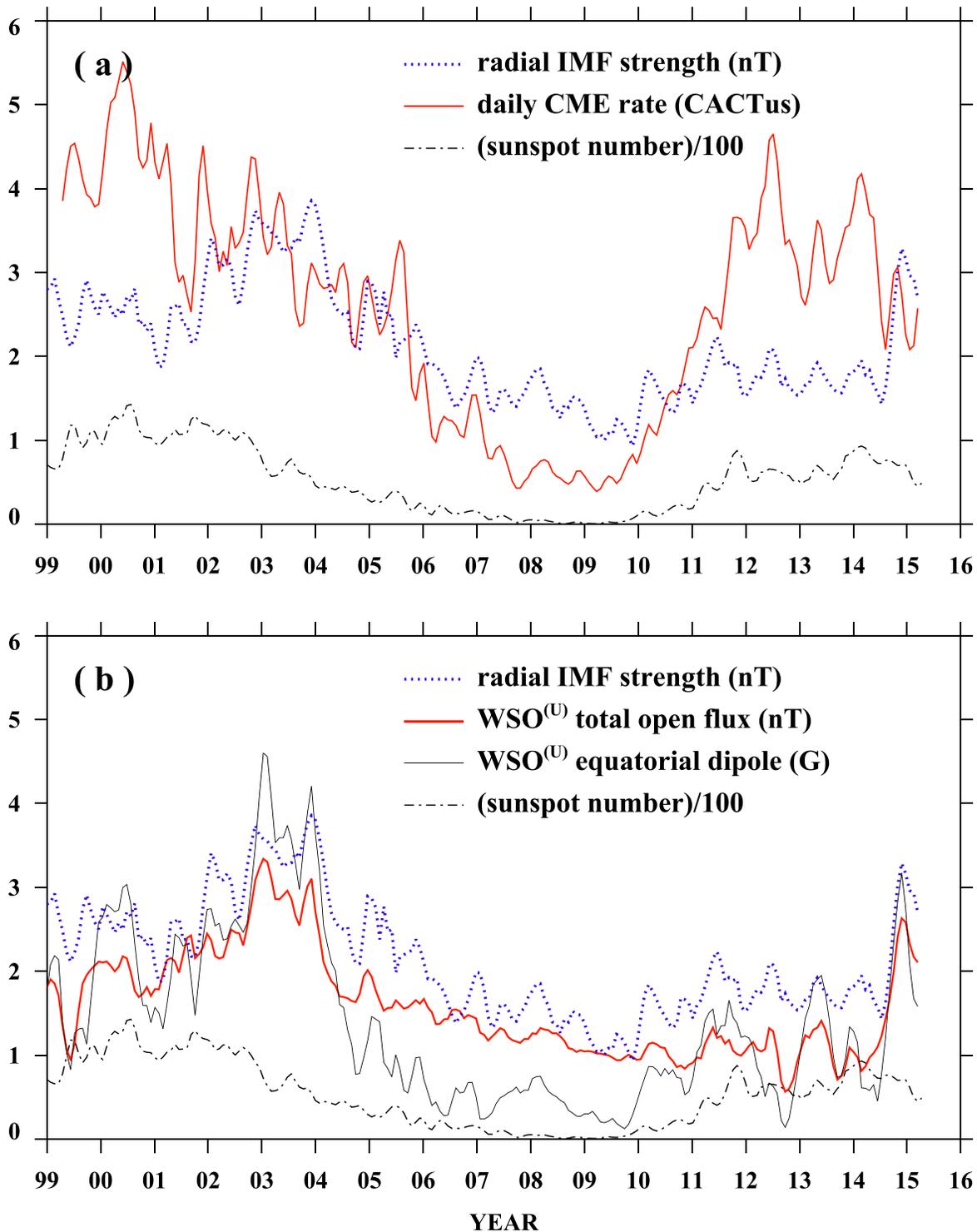

**Figure 1.** (a) Variation of the CME rate and radial IMF strength during 1999–2015. Here and in subsequent figures, all curves represent 3-CR running means. Solid curve: $N_{\rm CME}$, the daily number of CMEs, as given by the CACTus catalog. Data from LASCO and COR2A/B have been combined as described in the text. Dotted curve: near-Earth radial IMF strength $B_E^{\rm obs}$ (nT), extracted in the form of daily averages from the OMNI database. Dashed–dotted curve: monthly mean sunspot number $R_{\rm I}$ ($\times 0.01$). $N_{\rm CME}$ varies in phase with $R_{\rm I}$ (cc = 0.89), but $B_E^{\rm obs}$ peaks just after sunspot maximum. (b) Variation of the Sun's total open flux and equatorial dipole strength during 1999–2015, derived from WSO photospheric field measurements multiplied by the Ulrich calibration factor ($4.5 - 2.5 \sin^2 L$). Thick solid curve: $B_E^{\rm WSO}$ (nT), the total open flux converted into a field strength at 1 AU. Thin solid curve: $D_{\rm eq}$ (G), the equatorial dipole component of the photospheric field. Dotted curve: $B_E^{\rm obs}$ (nT). Dashed–dotted curve: $R_{\rm I}$. $B_E^{\rm WSO}$ varies in phase with $B_E^{\rm obs}$ (cc = 0.90), with both showing large peaks in 2003 and late 2014.

To obtain a more realistic estimate, we make use of the ICME catalog of Richardson & Cane (2010)[3] to identify the times of individual events that occurred between 1996 and the present, and assign to each ICME a radial field strength $|B_r|_{\rm ICME}$ extracted from the OMNI database. The average ICME contribution to $B_E$ over a given CR is then found by summing over all $N_{\rm ICME}$ events that occurred during that 27.3 day interval:

$$B_E^{\rm ICME} \sim \sum_{i=1}^{N_{\rm ICME}} |B_r(i)|_{\rm ICME} \left[ \frac{\tau_{\rm ICME}(i)}{\tau_{\rm CR}} \right]. \quad (3)$$

---

[3] See http://www.srl.caltech.edu/ACE/ASC/DATA/level3/icmetable2.htm.

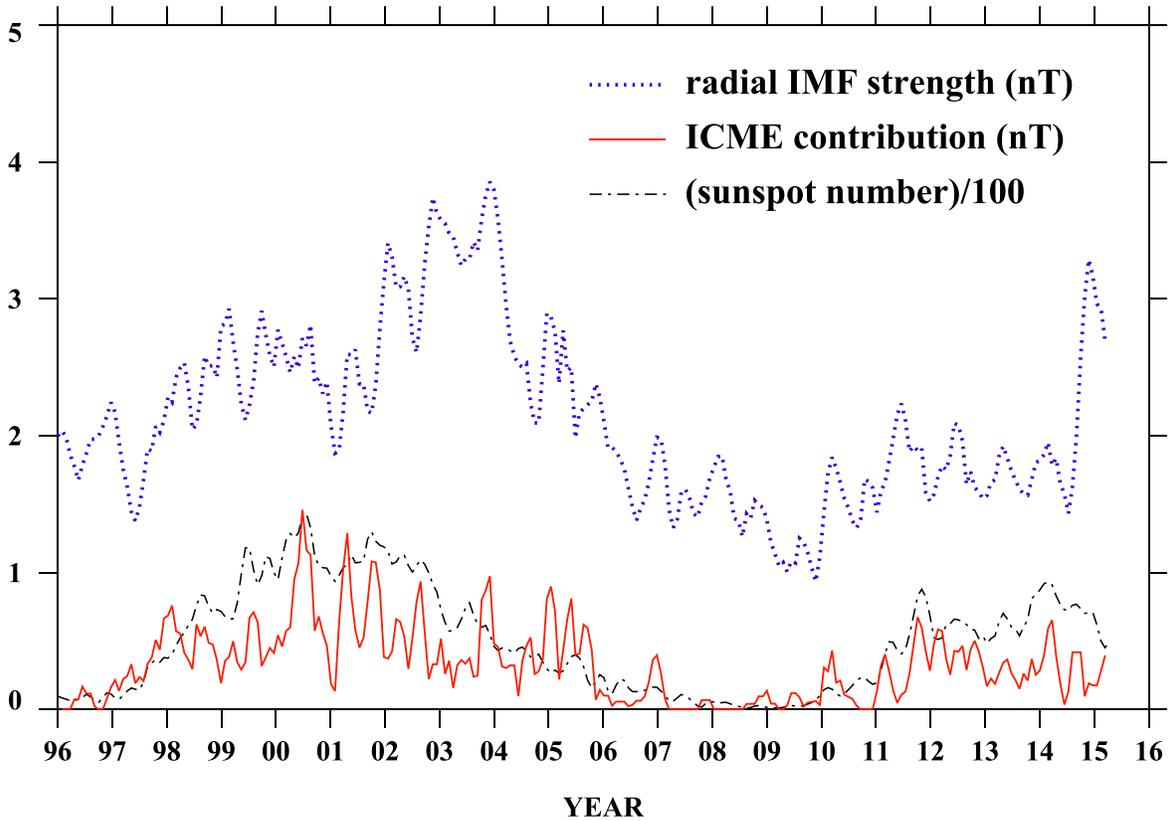

**Figure 2.** Variation of the radial field strength contributed by near-Earth ICMEs, $B_E^{\rm ICME}$ (nT) (solid curve), derived by summing the maximum radial fields associated with the individual events listed in the Richardson–Cane catalog. Also plotted are the OMNI measurements of $B_E^{\rm obs}$ (nT) (dotted curve), and $R_I$ (dashed–dotted curve). Even during the cycle 23 activity maximum (1999–2002), ICMEs contributed an average of only ∼23% to the observed near-Earth radial IMF strength. It is also clear that ICMEs cannot account for the large peaks in $B_E^{\rm obs}$ that occurred in 2003 and at the end of 2014.

Here, $|B_r(i)|_{\rm ICME}$ will be set to the largest daily averaged value of $|B_r|$ measured during the passage of ICME $i$. Following Riley (2007), $\tau_{\rm CME}(i)$ will be assigned a fixed value of 1.5 days; similar results are obtained if the observed duration of each event is used instead. For the 457 ICMEs listed in the Richardson–Cane catalog between 1996 May and 2015 April, the (largest daily averaged) radial field strengths and duration times have mean values of $\langle |B_r|_{\rm ICME} \rangle = 3.4$ nT and $\langle \tau_{\rm CME} \rangle = 1.1$ days, respectively. The ICME fields were strongest during the post-maximum years 2002–2004, when the average values of $|B_r|_{\rm ICME}$ were as high as ∼4.5 nT.

Figure 2 shows the time variation of $B_E^{\rm ICME}$ obtained from Equation (3), along with the observed variation of the radial IMF strength $B_E^{\rm obs}$. Even at sunspot maximum, ICMEs are evidently not the main contributor to the near-Earth radial IMF strength. Thus, when averaged over the interval 1999–2002, the ratio $B_E^{\rm ICME}/B_E^{\rm obs}$ has a value of $f_{\rm ICME} \simeq 0.23$; averaging over 2011–2014, we find an even smaller ICME contribution of $f_{\rm ICME} \simeq 0.18$. The ICME contribution peaked during 2000–2001, when $f_{\rm ICME}$ had an average value of 0.30. These results differ from those of Riley (2007) because the field strength of ∼8 nT that he assigned to each ICME is more than twice the maximum radial field strength in a typical ICME; only 5% of the ICMEs in the Richardson–Cane catalog have $|B_r|_{\rm ICME} \geqslant 8$ nT.

As was already suggested by the solar cycle variation of the global CME rate plotted in Figure 1(a), Figure 2 confirms that near-Earth ICMEs cannot account for the large post-maximum peaks in the radial IMF strength. This is made especially clear in late 2014, when the sudden doubling of $B_E^{\rm obs}$ occurred as $B_E^{\rm ICME}$ and $R_I$ were declining from their maxima in early 2014 (see also Sheeley & Wang 2015). Although the spike in $B_E^{\rm obs}$ and $B_E^{\rm ICME}$ seen near the end of 2003 reflects the well-known October–November "Halloween" events, the broad peak in the IMF strength centered on 2003 is cotemporal with the presence of large, 27-day recurrent high-speed streams in the ecliptic. These long-lived streams in turn arose as a consequence of the strengthening of Sun's equatorial dipole component, which was accompanied by the formation of large low-latitude coronal holes and equatorward extensions of the polar holes.

In Figure 3(a), we plot the WSO-derived radial field strength $B_E^{\rm WSO}$ during 1976–2015, the sum of the WSO and ICME fluxes $B_E^{\rm (WSO+ICME)} = B_E^{\rm WSO} + B_E^{\rm ICME}$ (from 1996 onward), and the measured radial IMF strength $B_E^{\rm obs}$. As expected, adding the ICME contribution to $B_E^{\rm WSO}$ boosts the level of the Sun's total open flux around the cycle 23 and 24 sunspot maxima. As a result, the correlation between the predicted and observed values of the radial IMF strength during 1996–2015 increases from $cc(B_E^{WSO}, B_E^{obs}) = 0.81$ to $cc(B_E^{(WSO+ICME)}, B_E^{obs}) = 0.85$.

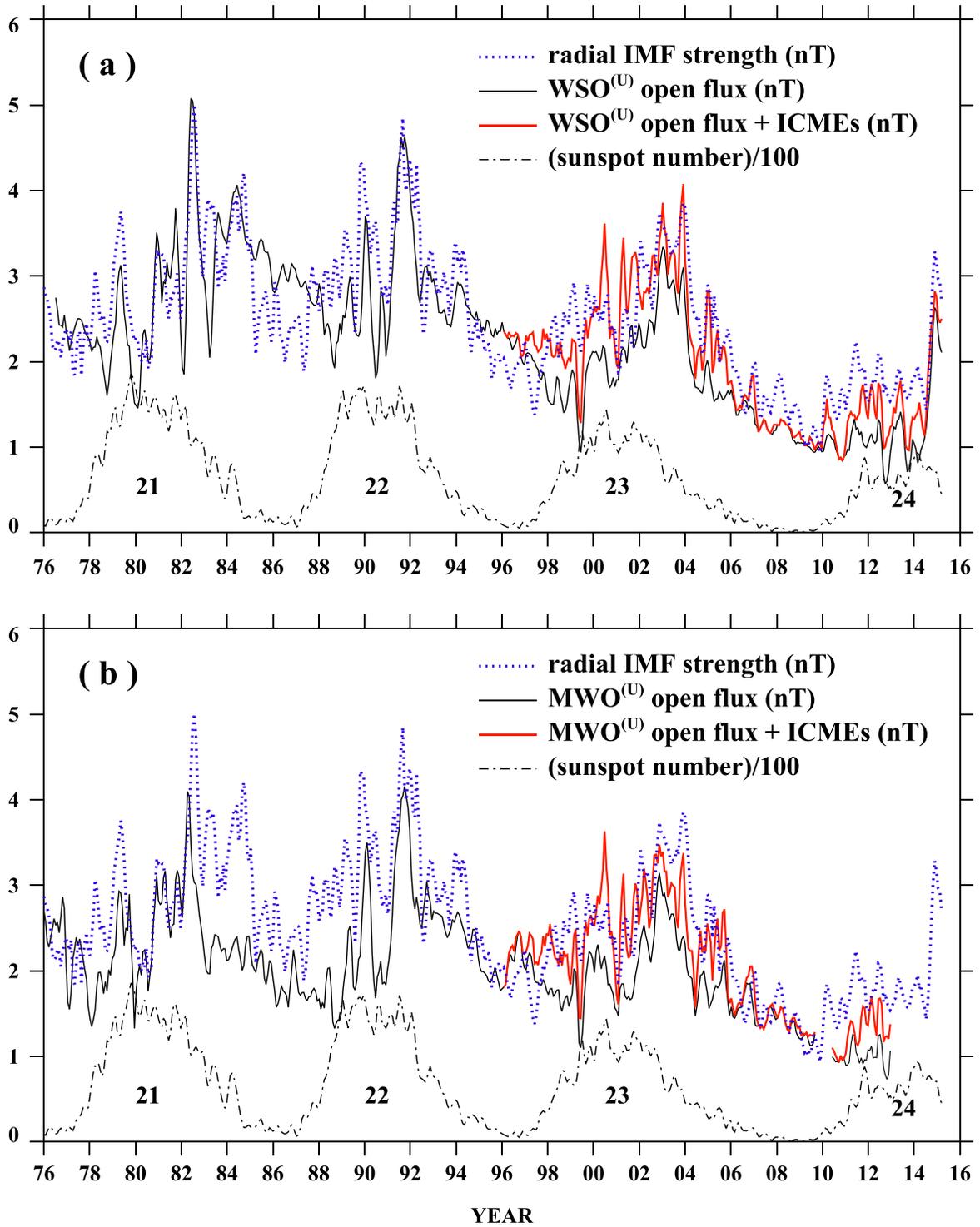

**Figure 3.** Effect of adding the contribution of ICMEs to the PFSS-derived open flux. (a) Black solid curve: $B_E^{WSO}$ (nT), the radial field strength predicted by applying a PFSS extrapolation (with $R_{ss} = 2.5\,R_\odot$) to WSO magnetograph measurements corrected by the factor $(4.5-2.5\sin^2 L)$. Red solid curve: $B_E^{(WSO+ICME)}$ (nT), the sum of $B_E^{WSO}$ and $B_E^{ICME}$. Dotted curve: observed radial IMF strength $B_E^{obs}$ (nT). Dashed–dotted curve: monthly mean sunspot number $R_I$ ($\times 0.01$). Including the ICME contribution improves the agreement between the predicted and observed IMF strength around sunspot maximum, with the correlation coefficient over the interval 1996–2015 increasing from 0.81 to 0.85. (b) Same as (a), except that the photospheric field measurements are from MWO instead of WSO.

In Figure 3(b), we have replaced $B_E^{WSO}$ by $B_E^{MWO}$, the open flux obtained from a PFSS extrapolation of MWO photospheric field measurements, again scaled upward by the Ulrich calibration factor $(4.5-2.5\sin^2 L)$. As in the case of the WSO open flux, adding the ICME contribution to the MWO open flux improves the agreement between the predicted and observed radial IMF strength, with the correlation during 1996–2015 increasing from $cc(B_E^{MWO}, B_E^{obs}) = 0.73$ to $cc(B_E^{(MWO+ICME)}, B_E^{obs}) = 0.80$.

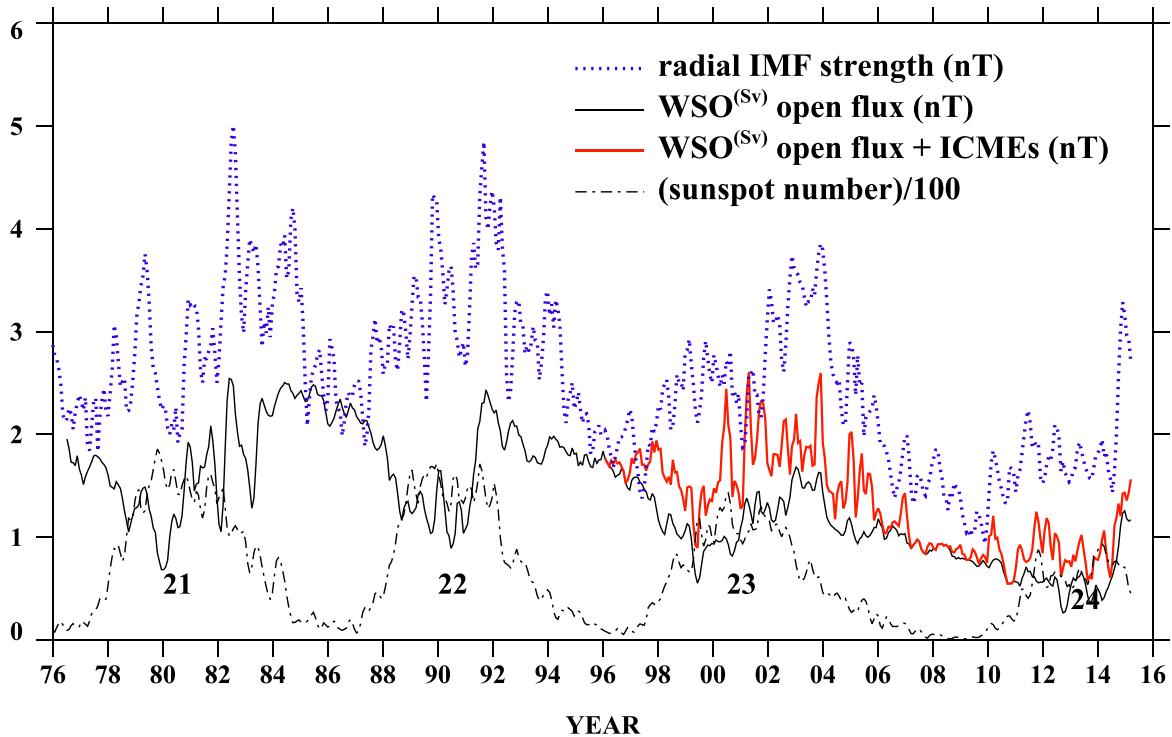

**Figure 4.** Same as in Figure 3(a), except that the WSO magnetograph measurements have been scaled upward by a factor of 1.8 (following Svalgaard et al. 1978) instead of by $(4.5-2.5\sin^2 L)$. The predicted radial IMF strength remains a factor of ~2 too low even after ICMEs are included (red solid curve).

Figure 4 shows the variation of $B_E^{WSO}$ and $B_E^{(WSO+ICME)}$ as in Figure 3(a), but here the WSO measurements have been multiplied by 1.8 (as advocated by Svalgaard et al. 1978) rather than by $(4.5-2.5\sin^2 L)$. As expected, the predicted values of the radial IMF strength remain a factor of ~2 too low even after the contribution of ICMEs is included.

### 4. CONCLUSIONS

The objective of this Letter has been to understand better the sources of the solar cycle variation of the IMF. Our results may be summarized as follows.

1. The variation of the radial (and total) IMF strength is out of phase with that of the total CME rate, with the latter closely following the sunspot number, but the IMF strength tending to peak just after sunspot maximum (Figure 1(a)).
2. The largest peaks in $B_E^{obs}$ (occurring in 1982, 1991, 2003, and late 2014) coincided with strong enhancements of the Sun's equatorial dipole component. Such enhancements are associated with the emergence of large active regions whose east–west dipole moments are in phase with that of the background field.
3. By summing over the radial field contributions of the individual events listed in the Richardson–Cane catalog between 1976 May and 2015 April, we find that ICMEs account for only a relatively small fraction of the radial IMF strength at Earth, even during sunspot maximum (Figure 2). Averaged over the interval 1999–2002 (2011–2014), $f_{ICME} = B_E^{ICME}/B_E^{obs} \sim 0.23$ (0.18).
4. Thus ICMEs cannot account for the observed factor of ~2 variation of the IMF strength over the solar cycle, which must have its origin in the long-term evolution of the Sun's large-scale field, in particular of its total dipole strength.
5. When the contribution of ICMEs is added to the WSO open flux, scaled upward by the Ulrich correction factor $(4.5-2.5\sin^2 L)$, the correlation between the predicted and observed variation of the radial IMF strength during 1996–2015 increases from 0.81 to 0.85 (see Figure 3(a)).
6. If the WSO open flux is instead scaled upward by the Svalgaard et al. saturation correction of 1.8, the predicted radial IMF strength remains a factor of ~2 too low even after ICMEs are included (Figure 4).

Our conclusion that ICMEs are not the main contributor to the IMF strength or its solar cycle variation is consistent with the results of Richardson & Cane (2012) and Yermolaev et al. (2009). Richardson & Cane (2012; see also Richardson et al. 2002) divided the near-Earth solar wind into three flow types (high-speed streams, slow interstream wind, and CME-related flows), and found that the average field at 1 AU is not dominated by the contribution of transients, even during high solar activity. Yermolaev et al. (2009) reached a similar conclusion using a slightly different classification scheme.

While our results provide support for the calibration correction of $(4.5-2.5\sin^2 L)$ derived by Ulrich (1992) at MWO over the factor of 1.8 found by Svalgaard et al. (1978) using the low-resolution WSO magnetograph, the reasons for this large discrepancy have yet to be clarified. A recalibration of the WSO instrument by comparing measurements in Fe I 525.0 nm and a nonsaturating line such as Fe I 523.3 nm, as was done by Ulrich (1992) and Ulrich et al. (2002, 2009) using the MWO magnetograph, would be highly worthwhile.

Our basic conclusion is that the solar cycle variation of the IMF mainly reflects the evolution of the Sun's lowest-order ($l = 1, 2$) multipoles, with CMEs giving rise to short-term fluctuations but providing only a rather modest contribution to the average IMF strength at sunspot maximum. The "floor" in the IMF is determined by the Sun's axial dipole strength at sunspot minimum, which may vary from cycle to cycle and which was probably much lower during the Maunder Minimum than during 2008–2009.

We are greatly indebted to J. W. Harvey, I. G. Richardson, D. G. Socker, and R. K. Ulrich for helpful discussions. This work was supported by CNR.